\documentclass[twocolumn,floatfix]{aastex62}

\usepackage{savesym}
\savesymbol{tablenum}
\usepackage{siunitx}
\usepackage{bm}
\usepackage{mathtools}

\usepackage{color}
\definecolor{midblue}{RGB}{44,133,255}
\definecolor{navy}{RGB}{35,46,127}
\definecolor{tmp}{rgb}{0,.7,0.5}
\definecolor{cerulean}{rgb}{0.0, 0.48, 0.65}

\newcommand{\citeg}[1]{\citep[e.g.,][]{#1}}
\newcommand{\citesee}[1]{\citep[see][Paper I]{#1}}

\received{\today}
\revised{XXXX}
\accepted{XXXX}

\submitjournal{ApJ}

\shorttitle{sBH Migration in AGN Disks II}
\shortauthors{Secunda et al.}

\begin{document}

\title{Orbital Migration of Interacting Stellar Mass Black Holes in Disks around Supermassive Black Holes II. Spins and Incoming Objects}

\correspondingauthor{Amy Secunda}
\email{asecunda@princeton.edu}

\author{Amy Secunda}
\affil{Department of Astrophysics, American Museum of Natural History,
  Central Park West at 79th Street, New York, NY 10024, USA}
  \affil{Department of Astrophysical Sciences, Princeton University, Peyton Hall, Princeton, NJ 08544, USA}

\author{Jillian Bellovary}
\affiliation{Department of Astrophysics, American Museum of Natural
  History, Central Park West at 79th Street, New York, NY 10024, USA}
\affiliation{Department of Physics, Queensborough Community College, Bayside, NY 11364, USA}
\affiliation{Physics Program, The Graduate Center, City University of New York, New York, NY 10016, USA}

\author{Mordecai-Mark Mac Low}
\affiliation{Department of Astrophysics, American Museum of Natural
  History, Central Park West at 79th Street, New York, NY 10024, USA}
\affiliation{Center for Computational Astrophysics, Flatiron Institute, New York, NY, 10010, USA}

\author{K.E. Saavik Ford}
\affiliation{Department of Astrophysics, American Museum of Natural
  History, Central Park West at 79th Street, New York, NY 10024, USA}
\affiliation{Department of Science, Borough of Manhattan Community
  College, City University of New York, New York, NY 10007, USA}
\affiliation{Physics Program, The Graduate Center, City University of New York, New York, NY 10016, USA}
\affiliation{Center for Computational Astrophysics, Flatiron Institute, New York, NY, 10010, USA}

\author{Barry McKernan}
\affiliation{Department of Astrophysics, American Museum of Natural
  History, Central Park West at 79th Street, New York, NY 10024, USA}
\affiliation{Department of Science, Borough of Manhattan Community
  College, City University of New York, New York, NY 10007, USA}
\affiliation{Physics Program, The Graduate Center, City University of New York, New York, NY 10016, USA}
\affiliation{Center for Computational Astrophysics, Flatiron Institute, New York, NY, 10010, USA}

\author{Nathan W. C. Leigh}
\affiliation{Departamento de Astronom\'ia, Facultad de Ciencias F\'isicas y Matem\'aticas, Universidad de Concepci\'on, Concepci\'on, Chile}
\affiliation{Department of Astrophysics, American Museum of Natural History, Central Park West at 79th Street, New York, NY 10024, USA}

\author{Wladimir Lyra}
\affiliation{Department of Astronomy, New Mexico State University,
  Las Cruces NM 88003, 
  USA}

\author{Zsolt S\'andor}
\affiliation{Department of Astronomy, E\"otv\"os Lor\'and University, P\'azm\'any P\'eter s\'et\'any 1/A, H-1117 Budapest, Hungary}
\affiliation{Konkoly Observatory, Research Centre for Astronomy and Earth Sciences,
  H-1121 Budapest, Hungary}

\author{Jose I. Adorno}
\affiliation{Department of Astrophysics, American Museum of Natural
  History, Central Park West at 79th Street, New York, NY 10024, USA}
\affiliation{Department of Physics, Queens College, City University of New York, Queens, NY, 11367, USA}

\begin{abstract}
The masses, rates, and spins of merging stellar-mass binary black holes (BBHs) detected by aLIGO and Virgo provide challenges to traditional BBH formation and merger scenarios. An active galactic nucleus (AGN) disk provides a promising additional merger channel, because of the powerful influence of the gas that drives orbital evolution, makes encounters dissipative, and leads to migration. Previous work showed that stellar mass black holes (sBHs) in an AGN disk migrate to regions of the disk, known as migration traps, where positive and negative gas torques cancel out, leading to frequent BBH formation. Here we build on that work by simulating the evolution of additional sBHs that enter the inner disk by either migration or inclination reduction. We also examine whether the BBHs formed in our models have retrograde or prograde orbits around their centers of mass with respect to the disk, determining the orientation, relative to the disk, of the spin of the merged BBHs. Orbiters entering the inner disk form BBHs with sBHs on resonant orbits near the migration trap. When these sBHs reach $\gtrsim80$~M$_{\rm{\sun}}$, they form BBHs with sBHs in the migration trap, which over 10~Myr reach $\sim1000$~M$_{\rm{\sun}}$. We find 68\% of the BBHs in our simulation orbit in the retrograde direction, which implies BBHs in our merger channel will have small dimensionless aligned spins, $\chi_{\rm{eff}}$. Overall, our models produce BBHs that resemble both the majority of BBH mergers detected thus far (0.66--120~Gpc$^{-3}$~yr$^{-1}$) and two recent unusual detections, GW190412 ($\sim$0.3~Gpc$^{-3}$~yr$^{-1}$) and GW190521 ($\sim$0.1~Gpc$^{-3}$~yr$^{-1}$).
\end{abstract}


\section{Introduction}
\label{sec:intro}
The high rate of stellar mass binary black hole (BBH) mergers inferred from Advanced Laser Interferometer Gravitational Wave Observatory \citep[aLIGO;][]{aasi_2015} and Advanced Virgo \citep{Acernese_2014} detections, as well as the high masses and low spins of many of these mergers \citep{Abbott_2019}, has inspired much debate over which merger channels could best produce these mergers. Isolated binary evolution \citep{belczynski,belczynski_2010,postnov_2014,de_Mink_2016}, dynamical formation in globular clusters \citep[GCs;][]{benacquista_2013,Leigh_2014,wang_2016}, triple and quadruple systems \citep{Fragione_2020,Fragione_2019b,Fragione_2019a,Liu_2018}, young globular nuclear clusters \citep{Banerjee_2017,Antonini_2016}, and chemically homogeneous evolution in binaries \citep{Mandel_2016} are several potential merger channels. An additional merger channel proposed by \cite{Hopman:2006aa}, \cite{OLeary}, \cite{Antonini_rasio}, and \cite{rodriguez} suggests that over-massive stellar mass black holes (sBHs) are most likely to form in galactic nuclear star clusters. The apparent sBH cusp at the center of the Milky Way observed by \cite{Hailey_2018} lends further weight to this possibility. 

Here we focus on the scenario proposed by \cite{mckernan14,mckernan18}, who have suggested that the gas disks in active galactic nuclei (AGN) are especially favorable locations for the BBH mergers detectable by aLIGO and Virgo. \citet{mckernan14} point out that gas disks act to decrease the inclination of intersecting orbiters and harden existing binaries. Additionally, the orbiters in a gas disk exchange angular momentum with the surrounding gas leading to a change in the semi-major axes of their orbits, known as migration \citep{goldreich_tremaine}. As sBH orbiters in the disk migrate with speed dependent on their mass, they encounter each other and form BBHs, especially since sBHs orbiting in the same direction will encounter each other at relative velocities far smaller than in a gas-free star cluster \citep{mckernan2012,mckernan18,leigh}. 

If a gas disk is locally isothermal, the gas torques cause all isolated orbiters to migrate inward \citep{goldreich_tremaine,ward,tanaka_ward}. However, \cite{paardekooper_mellema} showed that in the more realistic case of a radiatively inefficient disk with an adiabatic midplane, for some values of the radial density and temperature gradient the torque from the disk can instead cause outward migration. \cite{paardekooper2010} used analytic arguments and numerical simulations to model the sign and strength of migration. They found that at the boundaries between regions of inward and outward migration the torques cancel out, leading to an orbit with zero net torque where the migration halts \citep{lyra}, referred to as a migration trap. \cite{bellovary} applied the \citet{paardekooper2010} migration torque model to two steady-state AGN disk models derived by \cite{Sirko:2003aa} and \cite{thompson} and showed that migration traps were present in both.

Building on this work, \citet[][hereafter Paper I]{secunda} used an augmented N-body code \citep{sandor,horn} that incorporates the \citet{paardekooper2010} migration torques (see however, Appendix \ref{A} for alternative torque models) to simulate the migration of ten sBHs in a \citet{Sirko:2003aa} AGN disk. Paper I found that migrating objects encountered each other at a rapid rate, causing 60--80\% of sBHs in their simulations to form BBHs over 10~Myr. From these simulations Paper I estimated an upper limit of the merger rate parameterized in \citet{mckernan18} of 72~Gpc$^{-3}$~yr$^{-1}$. The most massive models in Paper I produced over-massive sBHs of around 100~M$_{\rm{\sun}}$. \cite{Tagawa_2020} ran N-body simulations combined with a semi-analytical model of a \citet{thompson} AGN disk and found a sBH merger rate of $\sim$0.02 - 60~Gpc$^{-3}$~yr$^{-1}$, with binary formation due mainly to dynamical friction and dynamical interactions. The upper limits of these two papers are similar because both papers used a broadly similar disk model, and made similar assumptions about the lifetime of the disk. \citet{Tagawa_2020} is different from Paper I and this paper, though, in that they only simulated the outer regions of their AGN disk, far beyond the location of the migration trap for a \citet{thompson} disk. \cite{Yang_2019b} carried out Monte Carlo simulations of sBHs ground down into alignment with an AGN disk for a variety of supermassive black hole (SMBH) masses and accretion rates and found a BBH merger rate of $\sim$4~Gpc$^{-3}$~yr$^{-1}$. However, it is difficult to compare their rate with Paper I, because they did not include the sBHs initially co-planar with the disk when it formed.

In this paper, we use the simulation developed in Paper I of the inner disk region surrounding the migration trap of a \citet{Sirko:2003aa} disk to further investigate the potential of an AGN disk to provide a significant merger channel for BBH mergers detectable by aLIGO and Virgo. First, we examine the impact of sBHs whose inclinations are ground down by the disk until they are co-planar with the disk \citep{mckernan14,mckernan18}, as well as sBHs that are migrating inward from beyond 1000~AU (see \S \ref{sec:diff_runs:body_create} and \S \ref{sec:results:body_create}). In Paper I we found that BBH formation was halted after several hundred kiloyears. This end to BBH formation occurs once a sBH in the migration trap becomes massive enough to lock sBHs migrating towards the trap in resonant orbits before these sBHs can form BBHs with each other or the sBH in the migration trap. sBHs in our simulation become locked in resonant orbits when they exert a periodic gravitational influence on each other, which leaves them trapped in their orbit and protected against most perturbations to that orbit. \cite{sandor} found that in a proto-planetary disk adding migrating Mars-mass bodies helped disrupt resonant convoys and led to mergers and the formation of Super-Earths. Since the mass ratio of planets to stars in a proto-planetary disk is similar to the mass ratio of sBHs to the SMBH in an AGN disk, we wish to investigate if adding our own additional orbiters will produce similar effects. 

Second, we use our models from Paper I to predict what fraction of BBHs will orbit around their centers of mass in the prograde versus retrograde direction relative to the direction of rotation of the gas disk (see \S \ref{sec:diff_runs:pvr} and \S \ref{sec:results:provret}). {\citet{mckernan18,mckernan19} showed that the direction that BBHs orbit around their centers of mass affects the spin of the sBHs resulting from the mergers of these BBHs. The spins of these sBHs have implications for the value of the dimensionless aligned spin, $\chi_{\rm{eff}}$, of hierarchical mergers, which Paper I suggests are common in AGN disks. Given the low values of $\chi_{\rm{eff}}$ observed in many of the aLIGO detections thus far \citep{Abbott_2019}, predicting the spins of BBHs formed in our simulations} will help constrain the contribution of mergers of sBHs in AGN disks to aLIGO and Virgo detections.

The outline of this paper is as follows. In \S \ref{sec:methods} we describe the methods used in this paper, including a brief outline of the N-body code used (\S \ref{sec:n_bod}), and how the N-body code has been altered in order to simulate sBHs that migrate into the inner disk from beyond 1000~AU (\S \ref{sec:diff_runs:body_create}), and to examine the direction of the orbits of BBHs formed in our simulations (\S \ref{sec:diff_runs:pvr}). In \S \ref{sec:results} we discuss our results, including the the impact of additional sBHs migrating inward (\S \ref{sec:results:body_create}), and the orientations of the orbits of the BBHs in our simulations (\S \ref{sec:results:provret}). We discuss the implications and caveats of our findings in \S \ref{sec:discuss}.

\begin{figure}[htb!]
\label{fig:sirkoM}
\includegraphics[width=0.48\textwidth]{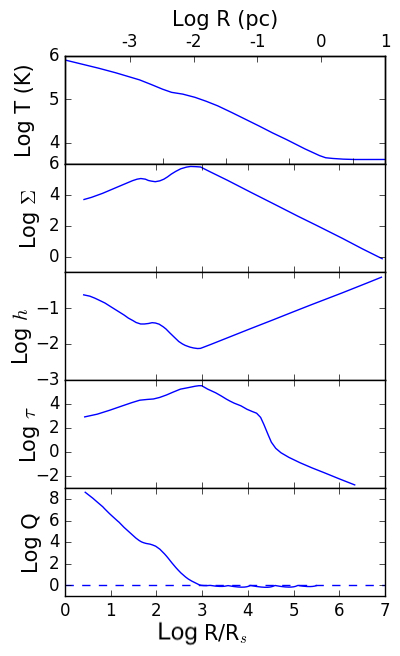}
\caption{\citet{Sirko:2003aa} SMBH accretion disk model used in our simulations. From top to bottom are plotted the midplane temperature $T$, surface density $\Sigma$ (in g~cm$^{-2}$), disk aspect ratio $h$ ($H/r$), optical depth $\tau$, and Toomre parameter $Q$ as a function of Schwarzschild radius $R_{\rm s}$. The top axis represents the translation from Schwarzschild radii to parsecs for a $10^8$~M$_{\sun}$ SMBH.}
\end{figure}

\section{Methods}
\label{sec:methods}
In this Section we begin by briefly outlining the set up of the N-body code used in Paper I (\S \ref{sec:n_bod}). Next we detail how we update our N-body code in order to first, simulate incoming sBHs that are either ground down into the disk or migrate inward from beyond 1000~AU (\S \ref{sec:diff_runs:body_create}) and second, determine whether the BBHs formed in our simulations orbit in the retrograde or prograde direction relative to the orbit of the gas disk (\S \ref{sec:diff_runs:pvr}). These updates allow us to further examine how efficient an AGN disk would be at producing the BBH mergers that have been detected by aLIGO and Virgo.

\subsection{N-Body Code}
\label{sec:n_bod}
We use the Bulirsch-Stoer N-body code described by \cite{sandor} that was modified by \cite{horn} to include a migration force modeled using the analytic prescription of \citet{paardekooper2010}, a dampening force from dynamical friction derived from the timescales for eccentricity and inclination dampening given by \cite{cresswell_nelson}, and a perturbation spectrum that models a turbulent force driven by magnetorotational instability following \cite{Laughlin_1994} and \cite{ogihara}. These forces depend on the local AGN gas disk properties, including temperature, surface density, opacity, and scale height, as well as the gradients of temperature and surface density. The functional form of these forces can be found in Paper I (Section 2.2 for the migration torques, Section 2.3 for the turbulent force, and Section 2.4 for the dampening forces). We also discuss our choice to use fully unsaturated static torques in our models in Appendix \ref{A}.

The initial conditions of the disk, including disk parameters, and the initial orbital parameters of the sBHs are identical to those of Paper I to allow for an easy comparison. The AGN gas disk properties in our models are taken from the \citet{Sirko:2003aa} AGN disk model, which is a modified Keplerian viscous disk model \citep{shakura_sunyaev}, with a high accretion rate fixed at Eddington ratio 0.5. We use a SMBH mass of ~$\textit{M}_{\rm{SMBH}}$ = $10^8$~M$_{\sun}$. The total mass of the disk integrated out to \num{2e5}~AU is $\num{3.7e7}$~M$_{\sun}$. The midplane temperature, surface density, scale height, optical depth, and Toomre Q as a function of radius in this model are plotted in Figure~\ref{fig:sirkoM}. These values all depend on SMBH mass through a series of 11 equations. Here we only study a $10^8$~M$_{\sun}$ SMBH, but we believe the qualitative behavior of our simulations would remain the same with a different SMBH mass (see \S \ref{sec:discuss}).

Our simulations neglect forces exerted by the sBHs on the gas disk aside from those implicitly modeled by the migration torques, the effects of accretion onto the SMBH and the sBHs, and general relativistic effects. We consider a BBH to have formed in our simulations when two conditions are met. The first is that two sBHs must be within a mutual Hill radius,
\begin{equation}
	\label{eq:rhill}
	R_{\rm mH} = \left(\frac{m_{\rm i} + m_{\rm j}}{3\textit{M}}_{\rm{SMBH}}\right)^{1/3} \left(\frac{r_{\rm i} + r_{\rm j}}{2}\right),
	\end{equation}
where $m_{\rm i}$ and $m_{\rm j}$ represent the masses of the two sBHs and $r_{ \rm i}$ and $r_{\rm j}$ represent their distances from the SMBH.

The second is that the relative kinetic energy of the binary,
	\begin{equation}
	\label{eq:KE}
    K_{\rm rel}=\frac{1}{2}\mu v_{\rm rel}^2,
	\end{equation}
where $\mu$ is the reduced mass of the binary, and $v_{\rm rel}$ is the relative velocity between the two sBHs, must be less than the binding energy at the Hill radius,
	\begin{equation}
	U=\frac{G m_{\rm i} m_{\rm j}}{2R_{\rm mH}}.
	\end{equation}
We refer to these criteria as the standard merger criteria. In our runs examining the ratio of prograde to retrograde binaries, the first standard criterion is tightened to 0.65~$R_{\rm{mH}}$ or 0.85~$R_{\rm{mH}}$ to help us examine the orbits of the BBHs (see \S \ref{sec:diff_runs:pvr}).

Interactions between BBHs and the gas disk within the Hill sphere are poorly understood \citeg{Miranda_2016,Tang_2017,Moody_2019,Munoz_2019,Munoz_2020}. For simplicity, we therefore consider a BBH merged as soon as it forms. As a result, in our simulations BBH formation and mergers between two sBHs are equivalent. In actuality, these BBHs would harden due to gas torques on a timescale that depends on the distribution of gas within the Hill sphere of the binary, and the complicated effects of accretion onto the BBH and the resulting feedback. \citet{baruteau2011} modeled the hardening of binaries in a gas disk. They found that it takes roughly 1000 (200) orbits of binary stars orbiting prograde (retrograde) around the binary's center of mass with respect to the disk to halve the binary's semi-major axis. If we assume that once the binary's semi-major axis has been halved 20 times the BBH will merge rapidly through gravitational radiation, within the inner 1000~AU of the disk the BBHs in our simulations would take at most a few hundred years to merge. In Paper I, Figure 13, we compare the merger times of prograde and retrograde orbiting BBHs with the migration rate of sBHs up to 50~M$_{\rm{\sun}}$ as a function of radial distance from the SMBH. Because the BBHs in our standard merger criteria are separated by $R_{\rm{mH}}$ when they form, the time until they merge depends only on their radial distance to the SMBH. Figure 13 in Paper I illustrates that the timescale over which BBHs merge is negligible compared to other dynamical timescales. In addition, in Paper I we also showed that a third migrator is unlikely to ionize a BBH in our simulation, which would prevent the merger from taking place. 

Nonetheless it is possible for orbiters to escape once they meet our standard merger criterion. We discuss this possibility in Section \ref{sec:results:provret}, where we actually evolve BBHs down to 0.65~R$_{\rm{mH}}$ in order to examine the direction of the BBH's orbit. Therefore, even if BBHs do not truly merge instantaneously, it is likely that they will shortly after formation, before they can be disrupted by other interactions. Finally, we ignore the role of kicks due to gravitational recoil at BBH merger. Such kicks will typically cause small perturbations of $\mathcal{O}(10^2\mbox{ km~s}^{-1})$ to orbital velocities of $\mathcal{O}(10^4\mbox{ km~s}^{-1}) (a/10^{3}~$R$_{\rm{g}})^{-1/2}$, where $a$ is the semi-major axis of the orbiter. Therefore, we anticipate that kicks should not significantly change our results, although they may cause prompt electromagnetic counterparts to the mergers discussed here \citep{mckernan19}.

Our runs each begin with ten sBHs, as suggested by \cite{antonini}, who used the distribution of S-star orbits around Sgr $\rm{A}^{\rm{\star}}$ to estimate that at least $10^{\rm{3}}$ sBHs reside within 0.1~pc of the SMBH. This estimate is consistent with the population of $\mathcal{O}$($10^{\rm{4}}$) sBHs within 1~pc of Sgr $\rm{A}^{\rm{\star}}$ inferred by \citet{Hailey_2018}. We assume that sBHs are evenly distributed throughout the disk and estimate that there should be roughly ten sBHs within 1000 AU ($\approx0.005$~pc). While our assumption for the number of sBHs is based off of observations of the MW center, which is currently experiencing a quiescent phase, it has been suggested that at earlier times the MW's SMBH did undergo an active phase. The mass of Sgr $\rm{A}^{\rm{\star}}$ is also roughly 50 times smaller than the SMBH we model here. However, \citet{Tagawa_2020} suggest that a more massive SMBH would in fact have a larger number of sBHs, making our estimate of 10 sBHs conservative.

We perform three fiducial runs, F10, F20, and F30, each with ten sBHs having uniform masses of 10, 20, and 30 M$_{\rm{\sun}}$ sBHs, respectively. These fiducial runs are common in N-body simulations of planet formation \citeg{Chambers_2001,Kominami_2002,horn}, because they control for any random effects by setting initial masses and separations of semi-major axes uniformly. In these runs the sBH closest to the SMBH has an initial semi-major axis of 500 AU, and each successive sBH is separated by 30 $R_{\rm{mH}}$ from the one before it. We choose to distribute the initial sBHs in our fiducial run like this for two reasons. First, we are most interested in studying the region directly surrounding the migration trap. By giving the innermost sBH an initial semi-major axis of 500~AU and separating the sBHs by 30 $R_{\rm{mH}}$, we ensure that all three of our fiducial models will have sBHs initialized on either side of the migration trap with respect to the SMBH no matter what the masses of the sBHs are. Second, separating our sBHs by 30~$R_{\rm{mH}}$ ensures that initially the impact of the sBHs on each other is minimal.

We then run four more realistic mass distributions, LMA, LMB, HMA, and HMB, with \cite{kroupa} initial mass functions and maximum masses of 15~M$_{\rm{\sun}}$ and 30~M$_{\rm{\sun}}$ for the LMA/LMB runs and HMA/HMB runs, respectively. We select masses from the \cite{kroupa} initial mass function, by drawing from a Pareto power law probability distribution of sBHs with a probability density 
\begin{equation}
\label{eq:imf}
p(x) = \frac{(\alpha-1) m_{\rm 0}^{\alpha -1}}{x^{\alpha}},
\end{equation}
where $\alpha = 2.3$, the scale factor $m_{\rm 0} = 5 M_{\rm \sun}$, and $x$ is a mass that is drawn from the distribution. In runs LMA and LMB if a mass generated from the Pareto distribution is greater than 15~M$_{\rm{\sun}}$ then a new one is generated to ensure all sBHs have masses of less than 15~M$_{\rm{\sun}}$. The same is done in the HMA and HMB runs for masses greater than 30~M$_{\rm{\sun}}$. The initial semi-major axes of the sBHs in these runs are selected randomly from a uniform distribution between 200 and 1000 AU. We choose an outer semi-major axis limit of 1000~AU because our aim is to focus on the inner disk in the vicinity of the migration trap \citep[see however,][on the prevalence of BBH mergers away from a trap]{mckernan19,mckernan_2020,Tagawa_2020}. We choose the inner semi-major axis limit to avoid the region closest to the SMBH, where graviational radiation will have a large effect on the behavior of orbiters.

For all models, the eccentricities and inclinations are chosen randomly from Gaussian distributions with means 0.05 and 0, respectively, and standard deviations 0.02 and 0.05 radians, respectively, as in Paper I and \citet{horn}. The absolute value of the inclination generated is used, and if the eccentricity selected is negative a new one is selected until it is positive. We only select from low initial inclinations and eccentricities because it is outside the scope of this paper to examine the effects of the gas disk on highly eccentric and inclined orbiters \citep[see however,][]{MacLeod_2020,FAbj_2020}. The initial phases and arguments of pericenter are randomized.

For runs LMA, LMB, HMA, and HMB, the distance between all sBHs is calculated using the values generated. If any two sBHs are within 10~AU new initial positions will be generated to prevent two sBHs from being within less than a couple $R_{\rm{mH}}$ initially. We set a minimum spacing in order to prevent two sBHs from immediately being in very close proximity due to random chance. All sBHs in all seven runs are initialized orbiting in the prograde direction around the SMBH. However, an isotropic distribution of orbits would suggest that roughly half of sBHs should initially be orbiting in the retrograde direction when the disk forms. We defer a more thorough investigation of retrograde orbiters to future work, because we expect the differences in orbital evolution between prograde and retrograde orbiters to be non-trivial.  For example, recent work by \citet{secunda_2020b} suggests that retrograde orbiters in the disk would experience a rapid increase in eccentricity and decrease in semi-major axis that would lead to a build up of orbiters very close to the SMBH, or even coalescence with the SMBH. Additionally \cite{FAbj_2020}, show that retrograde sBHs orbiting on inclined orbits with respect to the disk are unlikely to get captured by the disk, reducing the number of co-planar retrograde orbiters relative to prograde orbiters.

The simulations were run for 10 Myr, which is within the range of estimated lifetimes for an AGN disk \citep{haehnelt,king_nixon,schawinski}. The names and initial masses (or ranges of initial masses) of the sBHs for each of the seven runs are shown in the first two columns of Table \ref{table:body_create}.

\subsection{Incoming Black Holes}
\label{sec:diff_runs:body_create}
Our simulated disk extends only to $10^{3}$~AU. However, assuming somewhere around $10^3$ sBHs are distributed somewhat uniformly in the inner $\sim$ 0.1~pc of an AGN disk \citep{antonini}, additional sBHs should migrate inward from beyond 1000~AU over time. The migration rates of a 10~M$_{\odot}$ sBH with an initial semi-major axis around a few thousand AU is on the order of $10^6$~Myr. Therefore if there are roughly an additional ten sBHs with intial semi-major axes around a few thousand AU it is reasonable to expect an additional sBH to migrate inward from beyond 1000~AU every 100~kyr.

Additionally, sBHs on inclined orbits that intersect the gas disk will be ground down into the plane of the disk \citep{bartos_2017,mckernan18}. \citet{FAbj_2020} showed that, for a \citet{Sirko:2003aa} AGN disk, sBHs with small initial inclination angles ($<15^{\circ}$) are preferentially ground down into the disk. In addition, these sBHs with small initial inclinations lose orbital energy as they are ground down into the disk. As a result, sBHs with large initial semi-major axes ($\sim 10^{5}~$R$_{\rm{g}}$) typically end up $\sim 10^{3}$~R$_{\rm{g}}$ from the SMBH in $<1$~Myr for small initial inclination angles ($\leq 10^{\circ}$). In our simulated disk with a SMBH of $10^8$~M$_{\odot}$ 1~R$_{\rm{g}} \sim 1$~AU. Therefore, assuming roughly 10 sBHs with small initial inclination angles and large initial semi-major axes, once again gives a rate of one incoming sBH per 100~kyr.

Here, for simplicity we conservatively assume that sBH repopulation of the inner disk ($<10^{3}$~AU) happens at a rate 10~Myr$^{-1}$. More massive sBH or BBHs will preferentially migrate in towards the trap. However here we simply assume that a 10~M$_{\odot}$ sBH migrates inwards from $10^{3}$~AU every $100$~kyr starting at $t=100$~kyr and apply this to all runs from Paper I (see Table \ref{table:body_create} in \S \ref{sec:results:body_create}). The initial inclinations and eccentricities of these incoming orbiters are chosen randomly in the same manner as described in \S\ref{sec:n_bod}, because we are modeling them once they are already in line with the disk. The results of these runs are described in \S \ref{sec:results:body_create}.

\subsection{Prograde Versus Retrograde Binaries}
\label{sec:diff_runs:pvr}

Whether a BBH is on a prograde or retrograde orbit around its center of mass with respect to its orbit around the SMBH will have implications for the spin of the sBH produced by the BBH merging \citep{mckernan18,mckernan19}. Thus far a majority of aLIGO and Virgo detections have been of BBH mergers with low values for their dimensionless aligned spins, $\chi_{\rm{eff}}$ \citep{Abbott_2019}. Therefore, if this trend is confirmed by future aLIGO and Virgo detections, knowing the fraction of BBHs formed in our simulations that orbit in the prograde versus the retrograde direction will help determine if an AGN disk is a prominent channel for producing the mergers detected by aLIGO and Virgo.

In order to ascertain whether the BBHs formed in our simulations are on prograde or retrograde orbits about their centers of mass, we need to evolve BBHs formed in our simulations beyond the point where they approach each other within a mutual Hill radius, instead of considering them merged as soon as they form. However, evolving these BBHs beyond their formation is computationally expensive as they can begin to rotate around each other rapidly, greatly decreasing the time step of our simulations. Moreover, we wish only to examine the initial conditions of the BBH, because of our neglect of the interactions between the gas and the BBH within 1~$R_{\rm{mH}}$. After experimenting with the trade-off of information on the BBHs orbit and computation time we decided to change our first standard merger criterion (see \S \ref{sec:methods}) to require that two sBHs approach each other within 0.65~$R_{\rm{mH}}$. This change allows us to determine the direction of the orbit of BBHs for four out of our seven runs (F10, F20, F30, and HMA), by both examining by eye the direction of orbit for the BBHs, and using the conservation of angular momentum (see \S \ref{sec:results:provret}).

However, for three runs, LMA, LMB, and HMB, changing the merger distance to 0.65~$R_{\rm{mH}}$ led to two sBHs forming a close, very rapidly rotating BBH that greatly decreased the time step of that run. As a result, for these three runs we use a merger distance criterion of 0.85~$R_{\rm{mH}}$, which we found prevented this problem from occurring, and are forced to use only conservation of angular momentum to determine the orientation of the BBH's orbit (see \S \ref{sec:results:provret}). 

In all runs, our second merger criterion, that the relative kinetic energy of two sBHs must be less than their binding energy in order to form a BBH, remains the same. The results of these runs are described in \S \ref{sec:results:provret}.

\section{Results}
\label{sec:results}

\subsection{Incoming Black Holes}
\label{sec:results:body_create}
\begin{figure*}
\centering
\begin{tabular}{lr}
\includegraphics[width=0.5\textwidth]{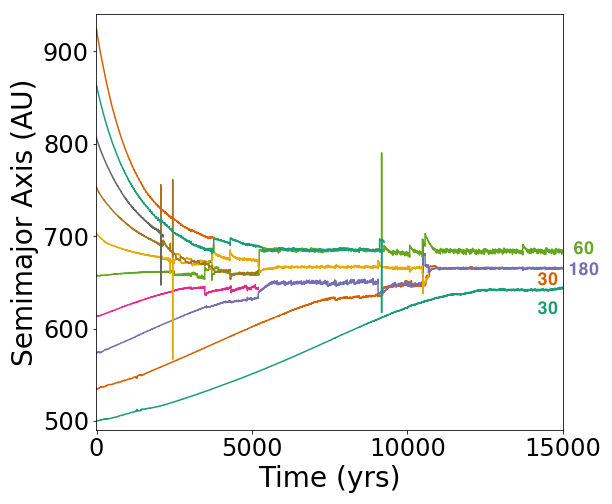} &
\includegraphics[width=0.5\textwidth]{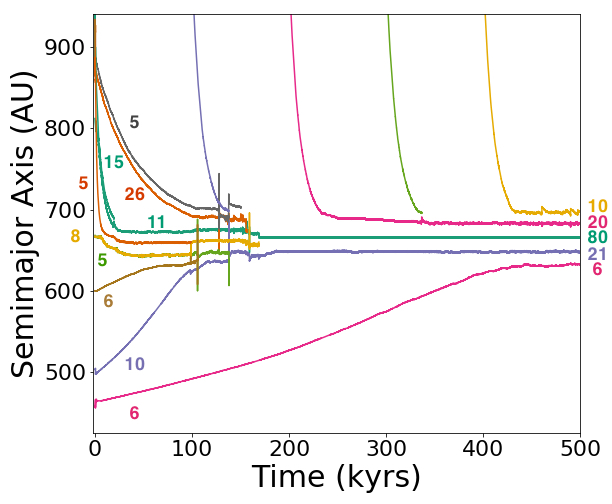}
\end{tabular}
\caption{The early migration histories of the F30 (left panel) and HMA (right panel) runs. Each colored line represents a different sBH, with the initial and final masses of that sBH (in M$_{\rm \sun}$) given in the same color on the left and right side of the figure, respectively. Initial masses in the F30 run (all 30~M$_{\rm \sun}$) and masses of incoming sBHs (all 10~M$_{\rm \sun}$) are not shown. Because the initial masses of the sBHs are greater in the F30 run, all sBHs have merged or are on stable orbits within 15~kyr, long before the first incoming black hole migrates in at 100~kyr. At 15 kyr in F30, the most massive sBH (purple) lies in the migration trap, which it shares with another lighter sBH (orange). Two orbiters (light and dark green) are in resonant orbits on either side.  In the HMA run, because the initial sBHs are less massive, and therefore take longer to migrate towards the migration trap, the first incoming sBH interacts with other orbiters before they have a chance to reach stable orbits. At 500~kyr the most massive sBH (dark green) orbits in the migration trap, surrounded by multiple resonant orbiters. The outer ones (pink and yellow) migrated in from beyond 1000~AU.}
\label{fig:early}
\end{figure*}

\begin{figure*}
\centering
\begin{tabular}{cc}
     & \includegraphics[width=0.95\textwidth]{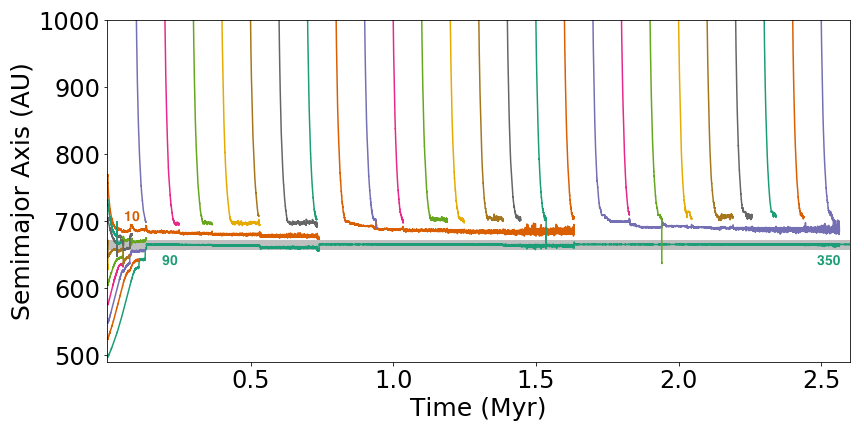} \\
     & \includegraphics[width=0.95\textwidth]{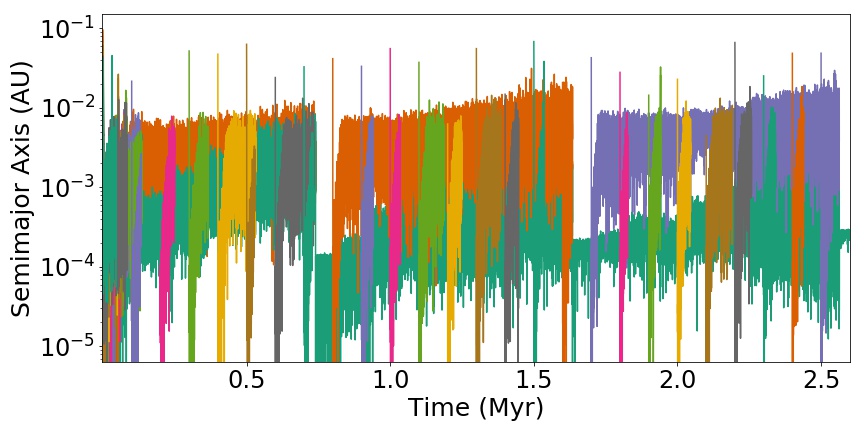}
\end{tabular}
\caption{The migration history of the first 2.6~Myr of the F10 run with sBHs migrating into the inner 1000~AU of the disk every 100~kyr and the same notation as Figure~\ref{fig:early}. The initial masses are not given in the top panel because they are all 10~M$_{\rm \sun}$. Instead we give the mass of the two remaining orbiters before the first merger involving an incoming sBH. The top panel shows the evolution of the semi-major axes of the sBHs as a function of time. The thick grey line shows the location of the migration trap at 665~AU. Most of the initial population of the inner 1000~AU of the disk form BBHs and merge within the first 100~kyr, forming a 90~M$_{\rm{\sun}}$ sBH that orbits in the migration trap around 665~AU (shown in dark green). Then as sBHs drift into the inner disk they form BBHs with the sBH that is on an outer resonant orbit. When the resonant orbiter becomes massive enough (about 80~M$_{\rm{\sun}}$) it merges with the sBH in the migration trap. This pattern repeats itself three times over the first 2.5~Myr. The bottom panel shows the evolution of the eccentricities of the sBHs as a function of time. The thin lines towards the top of the plot show the initial eccentricities of the sBHs migrating inward from the outer disk which are randomly assigned. These initial eccentricities are quickly dampened to under $10^{-3}$ until they are driven up to a few times $10^{-3}$ right before they form BBHs. The eccentricities of the sBHs on outer resonant orbits (in orange or purple) remain over $10^{-3}$ as their eccentricity is constantly being pumped by the additional orbiters migrating inward, with which they form additional BBHs. The dark green orbiter in the migration trap has its eccentricity dampened after each merger, and then gradually driven higher by the gravitational perturbations of the resonant orbiter as it becomes more and more massive until it reaches an eccentricity of over $10^{-3}$ and merges with the resonant orbiter.}
\label{fig:F10hist}
\end{figure*}

\begin{figure*}
\includegraphics[width=\textwidth]{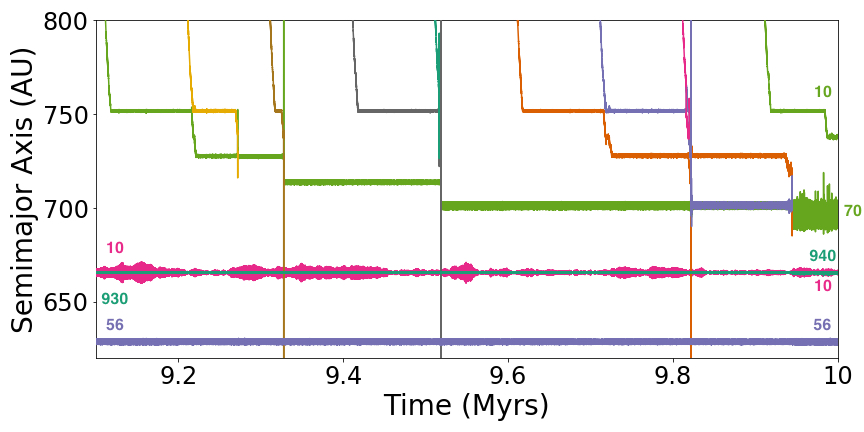}
\caption{The last megayear of the HMA run, with the same notation as Figure~\ref{fig:early}. Because the more massive sBH in the migration trap, shown in dark green, is 940~M$_{\rm{\sun}}$, it produces strong gravitational perturbations that make the migration history less periodic than in the F10 run shown in Figure \ref{fig:F10hist}. For example, multiple instances of BBH formation take place in rapid succession at just past 9.5~Myr when the incoming dark green, grey, and light green orbiters all merge, forming a 60~M$_{\rm{\sun}}$ sBH. sBHs also can end up on different orbits instead of merging. For example, just after 9.8~Myr, the purple sBH ends up in a trojan orbit with the light green sBH after interacting with the incoming pink and orange orbiters.}
\label{fig:HMA_late}
\end{figure*}

\begin{figure*}

\includegraphics[width=\textwidth]{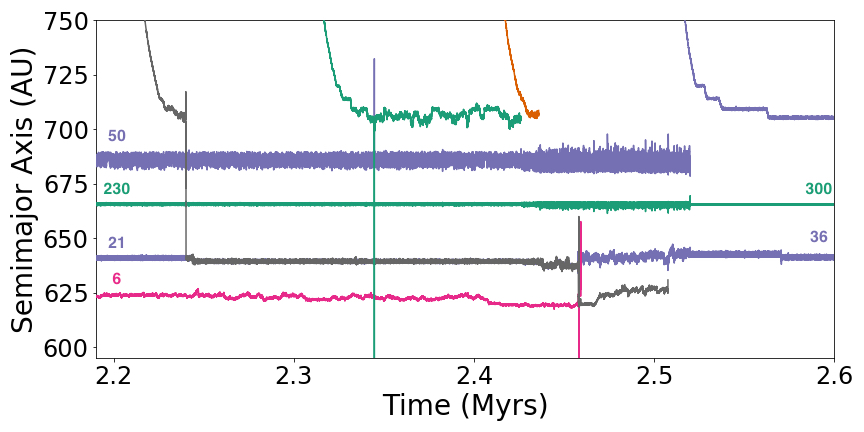}
\caption{The migration history for the HMA run from 2.2 to 2.6~Myr with sBHs migrating into the inner 1000~AU of the disk every 100~kyr and the same notation as Figure~\ref{fig:early}. The large number of sBHs on resonant and co-orbital orbits in this run leads to a rapid series of BBH formations set off by the sBHs migrating inward.}
\label{fig:HMA_res}
\end{figure*}

\begin{figure*}
\begin{tabular}{lr}
\centering
\includegraphics[width=0.5\textwidth]{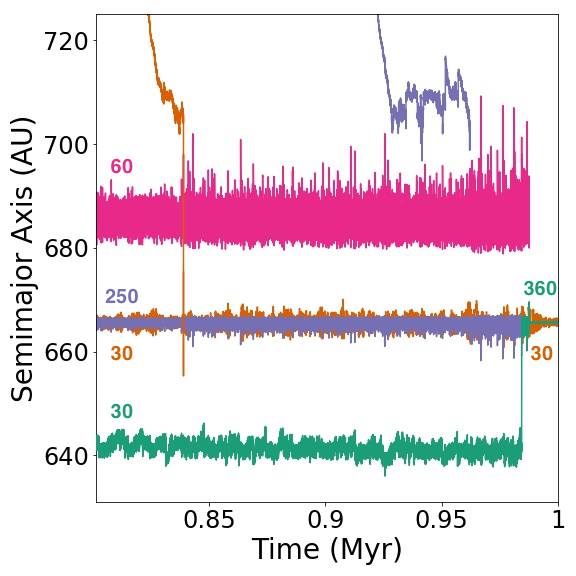} &
\includegraphics[width=0.5\textwidth]{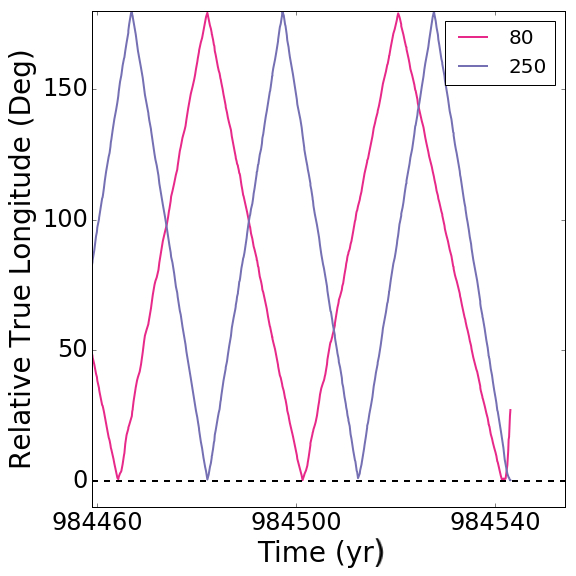}
\end{tabular}
\caption{An example of how incoming sBHs can destabilize orbits, leading to BBH formation, drawn from the F30 run with sBHs migrating into the inner 1000~AU of the disk every 100~kyr and the same notation as Figure~\ref{fig:early}. The left panel shows how the merger of an incoming sBH with the pink outer resonant orbit breaks the inner resonance of the green sBH. The resonance is broken when the relative true longitudes of the green sBH, pink sBH, and purple sBH in the migration trap approach zero. The right panel shows the true longitude of the pink and purple sBHs relative to the inner resonant orbiter shown in green.}
\label{fig:F30ex}
\end{figure*}

\begin{figure}
\centering
\includegraphics[width=0.5\textwidth]{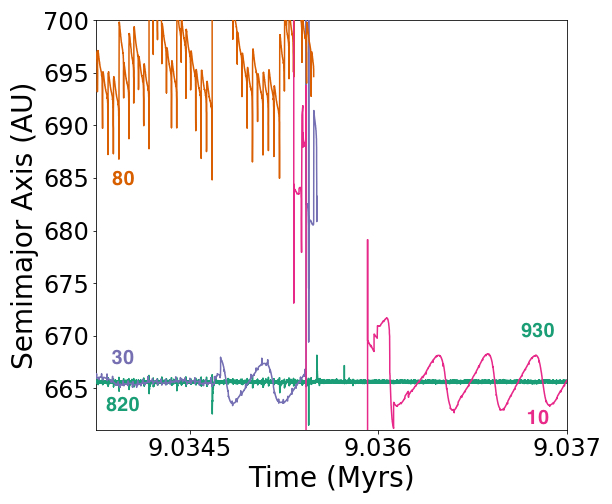}
\caption{An example of turbulence destabilizing orbits from the migration history of the HMA run with sBHs migrating into the inner 1000~AU of the disk every 100~kyr. This figure uses the same notation as Figure~\ref{fig:early}.
The turbulent mode leads to first the merger of the orange sBH with the purple sBH, which had previously been on a trojan orbit with the green sBH in the migration trap. Next, the purple black hole merges with the green black hole in the trap. The turbulent mode can be seen in the oscillations of the semi-major axes of the black holes.}
\label{fig:turb}
\end{figure}

\begin{figure}
    \centering
    \includegraphics[width=0.45\textwidth]{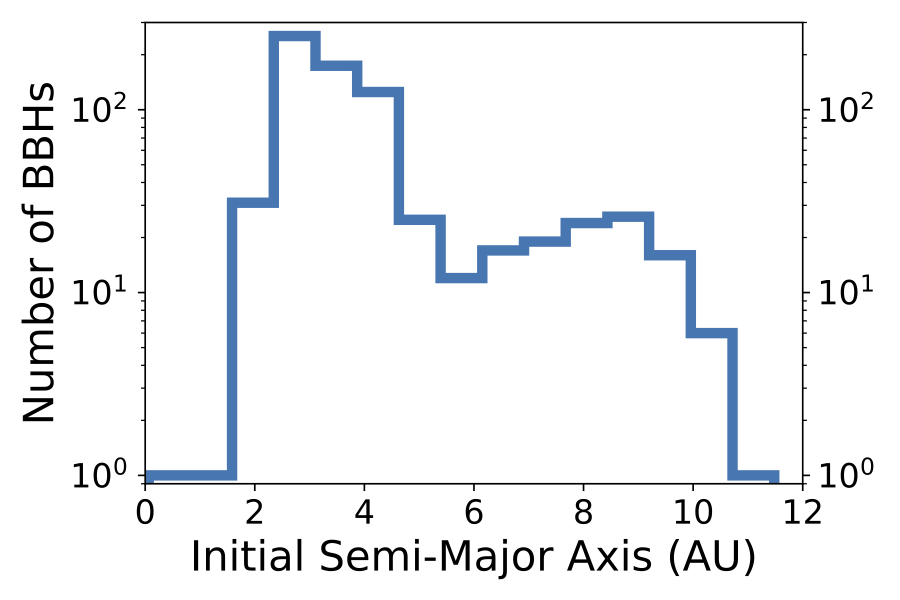} 
    \includegraphics[width=0.45\textwidth]{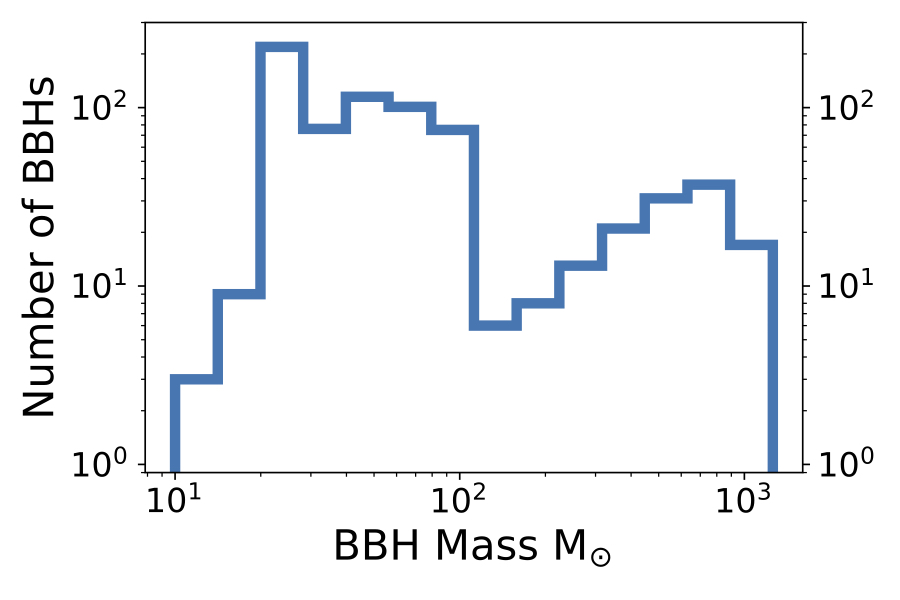}
    \includegraphics[width=0.45\textwidth]{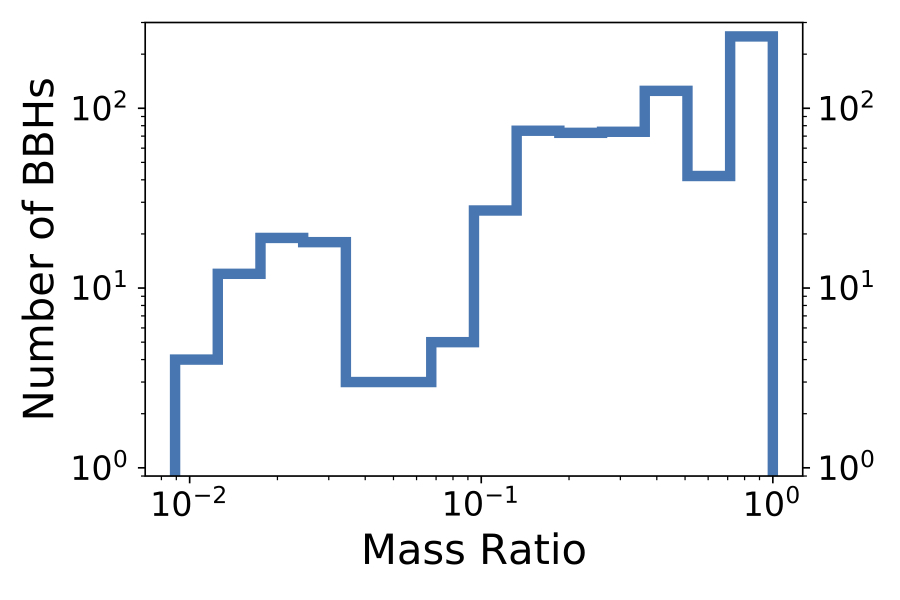}
    \caption{The distribution of initial semi-major axes in AU (top panel), BBH mass in solar masses (middle panel) and mass ratios (bottom panel) between the two black holes of BBHs formed in all of our runs combined. There is a peak around 3~AU from BBHs with total masses around 10-60~M$_{\rm \sun}$ whose member sBHs have mass ratios around 1. This peak extends to around 4.5~AU from BBHs with total masses up to around 90~M$_{\rm \sun}$ and mass ratios between 0.1 and 1. There is a smaller peak centered around 9~AU from around 6.5 to 10~AU from BBHs with total masses of around 350-1000~M$_{\rm{\sun}}$ with mass ratios also between 0.1 and 1.}
    \label{fig:smadist}
\end{figure}

\begin{figure}
    \centering
    \includegraphics[width=0.5\textwidth]{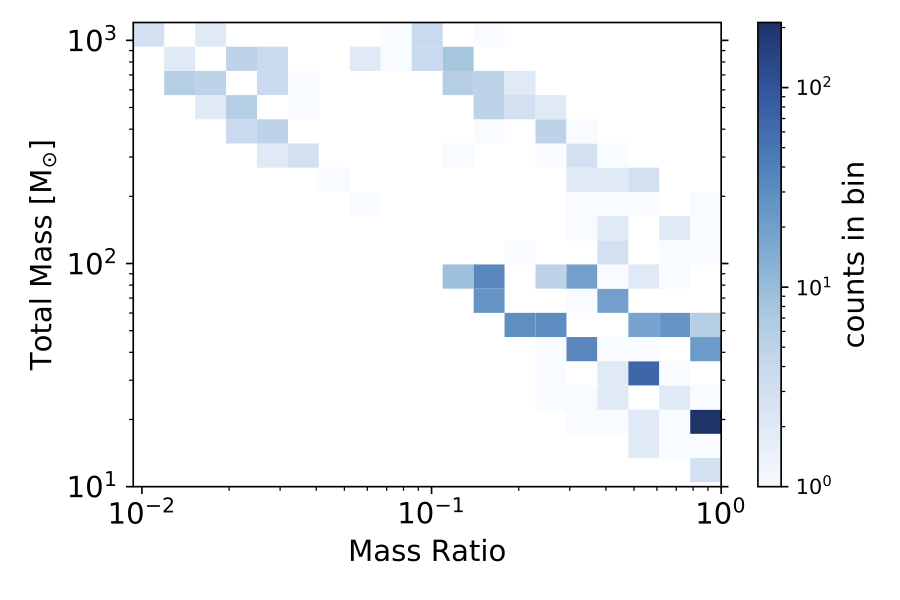}
    \caption{A 2D histogram of the mass ratios and total masses, in M$_{\rm \odot}$, of the BBHs formed in all of our runs combined. The overdensities at the right edge of the figure are of BBHs of nearly equal mass and total masses between around 20 and 60~M$_{\rm \sun}$ (class (1)). The two overdense diagonals on the lower right side of the figure are from uneven mass ratio BBHs with total masses between roughly 30 and 100~M$_{\rm \sun}$ (class (2)). The slight overdensity in the top right is from BBHs with similarly uneven mass ratios, but much larger mass from around 100--1000~M$_{\rm \sun}$ (class (3)). Finally, the slight overdensity in the top left corner represents highly uneven mass ratio BBHs, which tend to range from 300--1000~M$_{\rm \sun}$.}
    \label{fig:2Dhist}
\end{figure}

\begin{table}[t!]
	\centering
	\caption{Models with sBHs migrating in from beyond 1000~AU. Column 1: Name of run; Column 2: initial masses (or range of masses) of bodies in M$_{\rm \sun}$ before mergers and additional migrators begin drifting in; Column 3: the total combined mass of all bodies in the run in M$_{\rm \sun}$ including the 1000~M$_{\rm \sun}$ from the additional migrators; Column 4: the mass of the most massive sBH at the end of the run in M$_{\rm{\sun}}$; Column 5: the time each run takes to form a black hole over 100~M$_{\rm{\sun}}$ in kyr.}
	\label{table:body_create}
	\begin{tabular}{lcccr}
		\hline
		Run & $M_{\rm sBH}$ & $m_{\rm tot}$ & $m_{\rm max}$ & $T_{\rm form}$\\
		\hline
		F10 & 10 & 1100 & 960 & 750 \\
		F20 & 20 & 1200 & 1010 & 483 \\
        F30 & 30 & 1300 & 980 & 5.2 \\
        LMA & 5--15 & 1074 & 984 & 957 \\
        LMB & 5--15 & 1100 & 940 & 940 \\
        HMA & 5--30 & 1097 & 940 & 850 \\
        HMB & 5--30 & 1095 & 930 & 1100 \\
	\end{tabular}
\end{table}

Incoming sBHs migrating into the inner disk every 100~kyr can either represent sBHs migrating inward from the outer regions of the disk, or sBHs whose orbits have been ground down by dynamical friction from inclined orbits into alignment with the disk. The initial set-up of the runs described here (see \S \ref{sec:n_bod}) are identical to the runs performed in Paper I. Therefore the merger histories of each run is identical to the merger histories of the runs from Paper I up until the first incoming sBH migrates in from beyond 1000~AU at 100~kyr. The early migration history of the F30 and HMA runs are shown in the left and right panels, respectively, of Figure~\ref{fig:early}. More massive sBHs have a higher migration rate (see e.g. Figure 13 of Paper I). As a result, in the more massive runs, most if not all of the initial ten sBHs will have either formed BBHs or are on stable resonant orbits before the first additional orbiter migrates in. For example, the entire initial population of the F30 run, where all sBHs are initially 30~M$_{\rm{\sun}}$, are on stable orbits by 15~kyrs (see left panel of Figure~\ref{fig:early}). On the other hand, in the less massive runs, migrators drifting in will have an effect on the migration history of the initial ten sBHs. The right panel of Figure~\ref{fig:early} shows the first 500~kyr of the HMA run, where most sBHs are initially less than 10~M$_{\rm{\sun}}$. As a result, the first incoming sBH interacts with the initial orbiters before they reach stable orbits. In all runs, the most massive sBH ends up in the migration trap within the first few hundred kiloyears and from then on typically forms a BBH with all sBHs that are more than  $\sim80$~M$_{\rm \sun}$, making it often the only black hole $\gtrsim80$~M$_{\rm{\sun}}$.

Figure~\ref{fig:F10hist} shows the evolution of the semi-major axes (top panel) and eccentricities (bottom panel) of the orbiters for the first 2.6~Myr of the F10 run. After the initial few incoming sBHs reach the inner disk, the merger histories of these runs become somewhat periodic, repeating the following pattern. The sBH on the outermost resonant orbit will continuously form BBHs and merge with sBHs migrating inward from the outer disk. As this outermost resonant orbiter becomes more massive, its semi-major axis will start to oscillate as the collective mass between itself and the sBH in the migration trap becomes great enough for the resonant sBH to feel a strong gravitational perturbation. The semi-major axis of the sBH in the migration trap will typically oscillate less than the semi-major axis of the resonant orbiter, because the sBH in the trap is usually significantly more massive than the resonant orbiter, and due to the equipartition of energy, the oscillations are inversely proportional to mass. For example, in the top panel of Figure~\ref{fig:F10hist} the purple, 60-80~M$_{\rm{\sun}}$, resonant orbiter oscillates a lot more than the green, 270~M$_{\rm{\sun}}$ sBH in the migration trap, before they form a BBH at around 2.5~Myr.

When the outermost resonant orbiter builds up enough mass, in this example around 80~M$_{\rm{\sun}}$, the gravitational perturbation will be strong enough to perturb the orbiter out of resonance towards the migration trap where it will form a BBH with the sBH in the trap. The next sBH that migrates inward will end up on a resonant orbit until it too becomes massive enough to form a BBH with the sBH in the migration trap. 

The eccentricities shown in the bottom panel of Figure~\ref{fig:F10hist} similarly follow a periodic evolution. The initial eccentricities of the additional orbiters migrating in from beyond 1000~AU, which can be seen as the single lines extending up beyond $10^{-2}$, are quickly dampened by the gas by around three orders of magnitude. However, as these orbiters migrate inward they begin to feel gravitational perturbations from the outer resonant orbiter which drives their eccentricities back up to a few times $10^{-3}$ before the two orbiters form a BBH. These constant interactions with incoming orbiters keep the eccentricity of the outer resonant orbiter relatively high, also a few times $10^{-3}$. As the resonant orbiter becomes more massive the green sBH in the migration trap begins to be perturbed by the resonant orbiter, which also drives its eccentricity up to a few times $10^{-3}$. Finally, right before the resonant orbiter and the sBH in the trap form a BBH, the eccentricity of the resonant orbiter can reach as high as $\mathcal{O}$($10^{-2}$). 

The general pattern shown in Figure~\ref{fig:F10hist} and described above repeats itself over the course of the entire run somewhat consistently and is seen in all of the different runs. Occasionally, an incoming sBH may merge with another incoming sBH before it has time to merge with the resonant orbiter, for example the brown and yellow incoming sBHs at around 0.5~Myr and the green and grey sBHs at around 0.75~Myr. These mergers typically occur when either the resonant orbiter or the sBH in the migration trap become massive enough to produce strong gravitational perturbations that accelerate incoming sBHs towards the migration trap faster than they would typically migrate. When two incoming sBHs merge before either merges with the resonant orbiter, this merger can lead to multiple mergers in rapid succession, relative to the slower somewhat periodic mergers that typically occur, or dramatic changes in orbit. Both multiple mergers and changes in orbit can be seen in Figure~\ref{fig:HMA_late} which shows the last megayear of the migration history of the HMA run, when there are two black holes in the trap, the more massive of which is 940~M$_{\rm{\sun}}$. Because the black holes in the migration traps have greater masses at later times, multiple mergers are more common towards the end of our simulation runs.

The evolution of our various runs also becomes less periodic than the evolution shown in Figure~\ref{fig:F10hist} if many sBHs end up on resonant orbits. The migration history from 2.2 to 2.6~Myr of the HMA run, which at around 2.4~Myr has six resonant orbiters, is shown in Figure~\ref{fig:HMA_res}. When there are many resonant orbiters at once, they generate a strong gravitational perturbation when they approach the same true longitudes, that is line up with each other with respect to the SMBH. Much like a more massive resonant orbiter would, these aligned resonant sBHs accelerate incoming sBHs inward at a faster rate than they would typically migrate. The large number of resonant sBHs can also lead to a rapid sequence of mergers as multiple resonant orbits can be destabilized by a single merger event. For example, the orange sBH in Figure~\ref{fig:HMA_res} accelerates inward rapidly at around 2.42 Myr and forms a BBH with the purple outer resonant orbiter a relatively short time after the green orbiter that had migrated in before it. The purple outer resonant orbiter is now 70~M$_{\rm {\sun}}$. When the true longitudes of the outer resonant orbiter aligns with the other orbiters it is able to decouple the 36 and 10~M$_{\rm {\sun}}$ co-orbital sBHs orbiting at 640~AU. Destabilizing these co-orbital sBHs leads to two additional mergers of sBHs on inner resonant orbits, and finally the merger of the outer purple resonant orbiter and the green orbiter in the migration trap. These five mergers occur within roughly 100~kyr of each other.

Figure~\ref{fig:F30ex} shows another example of how an incoming sBH can break an inner resonance that had been stable for nearly 1~Myr, from the F30 run. During the initial evolution of the F30 run, before additional sBHs start migrating in (see left panel of Figure~\ref{fig:early}), two sBHs end up on a trojan orbit in the migration trap at around 11~kyr, with one sBH on an inner resonant orbit. This configuration is maintained until the epoch shown in Figure~\ref{fig:F30ex}, with the more massive sBH growing from 180~M$_{\rm {\sun}}$ to 250~M$_{\rm {\sun}}$ (purple). The 30~M$_{\rm {\sun}}$ sBH in the trap is shown in orange, and the 30~M$_{\rm {\sun}}$ sBH on the inner resonant orbit is shown in dark green. 

At about 1~Myr the inner resonance is broken. The left panel of Figure~\ref{fig:F30ex} shows that at this time there are still two trojan orbiters in the migration trap and the second sBH to have migrated into the disk, shown in pink, is on an outer resonant orbit. This outer resonant sBH has merged with all of the subsequent sBHs migrating in from outside the disk. These mergers give the pink sBH a mass of 70~M$_{\rm {\sun}}$ when it merges with a sBH migrating in, shown in purple. The pink sBH is now 80~M$_{\rm {\sun}}$, which makes it massive enough that when the differences between the true longitudes of the purple 250~M$_{\rm {\sun}}$ sBH in the trap, the true longitude of the green inner resonant orbiter, and its own true longitude approach zero, it gravitationally perturbs the green sBH off its resonant orbit towards the trap, where it merges with the purple sBH.

The difference in true longitude between the purple and green sBHs and the pink and green sBHs is shown in the right panel of Figure~\ref{fig:F30ex}. The difference in true longitude between the green and pink sBHs is just barely past zero when the purple and green sBHs align, causing the resonance between them to be broken and the green sBH to merge with the purple sBH in the migration trap. The resulting 280~M$_{\rm {\sun}}$ sBH, represented in green, also ends up merging with the pink sBH shortly after. These merged sBHs continue along in a trojan orbit with the less massive, orange, sBH in the migration trap.

While the incoming sBHs in these runs were the most efficient mechanism for breaking resonances in our simulations, an additional mechanism for breaking resonances, which was also seen in the simulations in Paper I, is turbulence. AGN disks are sufficiently ionized (certainly in the inner regions) that the magnetorotational instability will drive turbulence. We model this turbulence as described in Sect.~\ref{sec:n_bod}.
  An example of turbulence leading to the disruption of both a resonant and a trojan orbit is shown in in Figure~\ref{fig:turb}. In this figure when a turbulent mode opens up in the vicinity of the migration trap the orange 80~M$_{\rm{\sun}}$ resonant orbiter merges with the purple, 30~M$_{\rm{\sun}}$ sBH, which had previously been on a trojan orbit with the green black hole in the migration trap. Next the black hole shown in purple, which is now 110~M$_{\rm{\sun}}$, merges with the green black hole in the migration trap, forming a black hole of mass 930~M$_{\rm{\sun}}$. The oscillations in semi-major axis from the turbulent mode are clearly visible, for example in the purple orbiter before it leaves the migration trap, and the pink orbiter once it reaches the migration trap.

Figure~\ref{fig:smadist} shows the distribution of the initial semi-major axes in AU, the masses in solar masses, and the mass ratios between the two sBH components of BBHs of the BBHs formed in all seven of our simulations combined. Note that the x-axis of the bottom panel is inverted to ease comparison with the upper panels. Figure \ref{fig:2Dhist} is a 2D histogram of the mass ratios and total masses, in solar masses, of the BBHs formed in all seven simulations combined. Because in our simulations two sBHs must be within a mutual Hill radius of each other to form a BBH (see \S \ref{sec:n_bod}), the initial semi-major axes of BBHs are related to the total masses of the BBHs. The distributions in Figures~\ref{fig:smadist} and \ref{fig:2Dhist} show the three most common classes of BBH formation in our simulations:
\begin{enumerate}
\item BBHs from two roughly even mass $\sim$10 -- 30~M$_{\rm{\sun}}$ sBHs, which typically have initial semi-major axes of around 3~AU when they form near the migration trap. These BBHs often form early in the run from sBHs that are initially in the inner 1000~AU of the disk, or when a resonant orbiter merges with an incoming sBH for the first time. Chaotic mergers that are not part of the typical pattern described in this section can also contribute to this class, such as when two incoming sBHs form a BBH before they have the chance to merge with the resonant orbiter, or when two resonant orbiters merge. The highest peaks in the distributions in Figure~\ref{fig:smadist} and the dense bins along the right edge of Figure~\ref{fig:2Dhist} are from this first class of BBHs.

\item BBHs from incoming 10~M$_{\rm{\sun}}$ sBHs forming BBHs with resonant orbiters, which often grow to around 80~M$_{\rm{\sun}}$ before forming BBHs with the black holes in the migration trap. The BBHs in resonant orbits form around 680--700~AU from the SMBH, which means they have initial semi-major axes up to around 4.5~AU. These BBHs will have progressively more uneven mass ratios of around 0.5-0.1. This second class of BBHs produces the shoulder in the first peak in the top of Figure~\ref{fig:smadist} that extends to 4.5~AU, the shoulder in the middle panel that extends from 30-90~M$_{\rm{\sun}}$, and is partially responsible for the shoulder in the bottom panel that extends from roughly 0.1 to 0.5. In Figure~\ref{fig:2Dhist} these BBHs produce the two diagonal overdensities in the lower right corner.

\item Resonant orbiters merging with the black hole in the migration trap at 665~AU, which tend to grow in mass from 100--200~M$_{\rm{\sun}}$ at earlier times, to as massive as 1000~M$_{\rm{\sun}}$ at later times. As a result, at earlier times BBHs form with semi-major axes of around 6.5~AU and at later times with semi-major axes as large as 10~AU. This third class leads to the second maximum in the distribution in the top panel of Figure~\ref{fig:smadist} between 6.5--10~AU and the last peak in the middle panel around 750~M$_{\rm{\sun}}$. This third class also contributes to the shoulder in the bottom panel that extends from roughly 0.1 to 0.5, because the mass ratio of an incoming sBH to a resonant orbiter growing in mass is similar to that of the resonant orbiter to the black hole in the migration trap. This third class is apparent in the overdensity in the top right of Figure~\ref{fig:2Dhist}.
\end{enumerate}

The final peak around 0.02 in the bottom panel of Figure \ref{fig:smadist} originates primarily from a low mass inner resonant orbiter or a trojan orbiter of the black hole in the migration trap merging with the black hole in the migration trap. The rare instance when an incoming sBH merges directly with the black hole in the migration trap also contributes to this final peak. The overdensity in the top left corner of Figure~\ref{fig:2Dhist} corresponds to this final peak. While a BBH with that extreme a mass ratio has never been detected, this peak constitutes only $\sim$~7\% of the BBHs formed in our simulations, suggesting these would be less likely to be detected.

The last two columns of Table \ref{table:body_create} give the maximum masses formed and the time it takes to form a 100~M$_{\rm{\sun}}$ object for all seven runs. In our runs sBHs typically merge to form sBHs of over 100~M$_{\rm{\sun}}$ in no more than 1~Myr. These models, which simulate the realistic influx of sBHs being ground down into the disk and migrating inward from beyond 1000~AU, further suggest that over the course of an AGN disk lifetime, black holes of nearly 1000~M$_{\rm{\sun}}$ can form. These intermediate-mass black holes (IMBHs) can form because nearly all inwardly migrating sBHs form BBHs over these timescales.

\begin{figure*}
\begin{tabular}{ccc}
\includegraphics[width=0.33\textwidth]{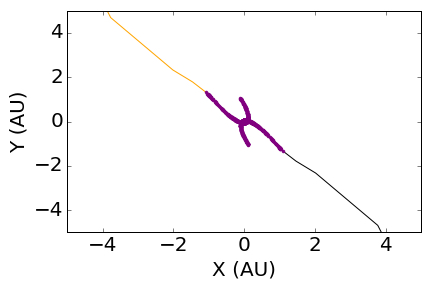} &
\includegraphics[width=0.33\textwidth]{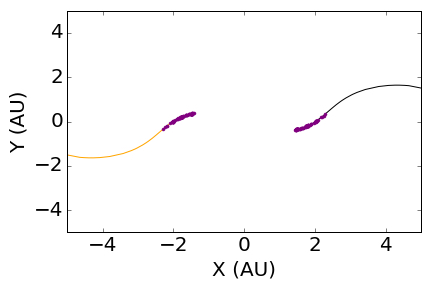}&
 \includegraphics[width=0.33\textwidth]{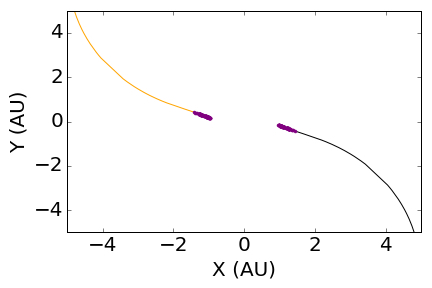}
\end{tabular}
\caption{The positions of two sBHs in a BBH with respect to their center of mass. The left panel shows an example of a BBH from the F20 run that is unambiguously orbiting in the prograde direction, the center panel shows an example of a BBH from the F30 run that is unambiguously orbiting in the retrograde direction, and the right panel shows an example of a BBH from the HMA run whose orbit is ambiguous. One sBH is plotted in orange, the other in black, until they are within 1~$R_{\rm{mH}}$ of each other, at which point they are both plotted in purple. Orbits are considered unambiguous when it is easy to tell from the points plotted in purple the direction that the two sBHs are orbiting around their centers of mass.}
\label{fig:direction}
\end{figure*}

\subsection{Prograde Versus Retrograde Binaries}
\label{sec:results:provret}

In our simulations described in \S \ref{sec:diff_runs:pvr}, most sBHs rapidly (relative to our adopted timestep) close the distance between 1~$R_{\rm{mH}}$, where we consider them bound, and 0.65~$R_{\rm{mH}}$, where we consider them merged. As a result, occasionally it can be difficult to determine by visual inspection whether the orbits around the centers of mass of sBHs in binaries are prograde or retrograde. The left and center panels of Figure~\ref{fig:direction} show two examples of BBHs unambiguously orbiting in the prograde and retrograde direction, respectively. The right panel of Figure~\ref{fig:direction} shows an example of an orbit that is more ambiguous. In all three panels the sBHs' orbits are plotted in orange or black until they are within 1~$R_{\rm{mH}}$ of each other and their binding energy is greater than their relative kinetic energy, i.e. our standard merger criteria, after which they are plotted in purple.

Since the orbits themselves are not always apparent, we use net angular momentum
to determine the direction of the BBHs' orbits. We apply this method to all BBHs in our simulations, even if the direction of their orbits is obvious from visual inspection. The sum of the angular momentum of the two sBHs, 
	\begin{equation}
	\bm{L}_{\rm{sum}}=\sum_i|m_{\rm{i}}\bm{v}_{\rm{i}}\times\bm{r}_{\rm{i}}|,
	\end{equation}
where $m_{\rm{i}}$, $\bm{v}_{\rm{i}}$, and $\bm{r}_{\rm{i}}$ are the mass, velocity, and distance from the SMBH of the two sBHs, respectively, is equal to $\bm{L}_{\rm{CM}} + \bm{L}_{\rm{bin}}$. $\bm{L}_{\rm{bin}}$ is the angular momentum of the orbit of the two sBHs in a binary around each other, and, 
	\begin{equation}
	\bm{L}_{\rm{CM}}=|m_{\rm{tot}}\bm{v}_{\rm{CM}}\times\bm{r}_{\rm{CM}}|,
	\end{equation}
where $m_{\rm{tot}}$ is the sum of the masses of the two bodies and $\bm{v}_{\rm{CM}}$ and $\bm{r}_{\rm{CM}}$ are the center of mass velocity and distance of the center of mass from the SMBH, respectively. Therefore if we subtract $\bm{L}_{\rm{CM}}$ from the sum of the angular momentum of the two sBHs, the difference will be the angular momentum of the orbit of the sBHs around each other. If the difference is positive, the binary must be orbiting in the prograde direction, and vice-versa for negative.

Our conservation of angular momentum method is limited by the fact that our N-body calculation includes a source of angular momentum from the migration torques that act on orbiters, causing them to migrate inwards and outwards. However, the differences in angular momentum are calculated for the first few timesteps after two sBHs meet the standard merger criteria, and the timescales of these timesteps are of order hours, whereas migration timescales are on the order of tens of thousands of years, for even the most massive sBHs in these models. Therefore angular momentum will be nearly conserved at least over several N-body timesteps after a BBH is first formed.

There were, however, five cases in which the difference $\bm{L}_{\rm{sum}} - \bm{L}_{\rm{CM}}$ changed sign between the time when two sBHs satisfied our standard merger criteria, and when they met the updated merger criteria (i.e.\ 0.65 or 0.85~$R_{\rm{mH}}$). We determined that in all 5 cases the change in $\bm{L}_{\rm{sum}} - \bm{L}_{\rm{CM}}$ is a result of a third body approaching close enough to either remove or contribute angular momentum to the binary. The amount of angular momentum exchanged is several orders of magnitude less than the angular momentum of any of the three orbiters, making it unlikely to dramatically harden or soften the BBH (see Paper I). Still, this exchange of angular momentum may be capable of affecting the evolution of the BBH. Four of the five cases where the sign of the difference changed were from runs where we could determine by visual inspection the direction of the orbit. In all four of these cases the direction corresponded to the initial sign of the difference, and so we label them accordingly. The fifth case was from the LMA run, where we only evolved BBHs down to a semi-major axis of 0.85~$R_{\rm{mH}}$, leaving us unable to determine the direction of the orbit by visual inspection. As a result we do not include that one BBH in Table \ref{table:pvr} or our analysis. We defer further study of the interactions between BBHs and other orbiters to future work that better resolves the evolution of a BBH interacting with gas within its Hill sphere.

For runs where the merger criterion was 0.65~$R_{\rm{mH}}$ we could compare our angular momentum calculation with plots of the orbits, if they were unambiguous. For all but two cases, there was agreement between our calculation and the plots. We have labeled the orbits of these two exceptions as ``Unknown." The number of BBHs orbiting around their center of mass in the prograde and retrograde directions, as well as the number of BBHs whose orbits were Unknown are shown in Table \ref{table:pvr}. The LMA, LMB, and HMB runs are shown in italics because these are the runs where we only trace the evolution of the sBHs down to a distance of 0.85~$R_{\rm{mH}}$, and so only angular momentum conservation is used to determine the direction of the BBH orbits. 

Overall our models found 15 BBHs orbiting in the prograde and 32 BBHs orbiting in the retrograde direction. The total number of BBHs we examined in our simulations is small, only 49, making it difficult to make any statistically significant statement. However, if it were equally likely for prograde and retrograde BBHs to form in an AGN disk the expectation value would be 24.5, which 32 exceeds by roughly 1.5$\sigma$. Additionally, while individually the difference between the number of prograde and retrograde BBHs in each run is not statistically significant, in most cases, it is still worth noting that in all runs, with the exception of the HMB run, there were more retrograde BBHs formed than prograde. This difference is most significant for the F30 run, which had 8 retrograde BBHs and 0 prograde. Therefore we can likely rule out that BBHs orbit primarily in the prograde direction around their centers of mass, and at least conservatively say that it is an even split between the number of BBHs orbiting in the prograde and retrograde direction.

Finally, with the exception of the few sBHs that formed a close, rapidly rotating BBH, changing the relative distance required for two sBHs to be merged from 1~$R_{\rm{mH}}$ to 0.85~$R_{\rm{mH}}$ or 0.65~$R_{\rm{mH}}$ had little qualitative effect on the migration history of our various runs. Therefore, our assumption that two sBHs will merge once they are within a mutual Hill radius of each other remains valid down to 0.65~$R_{\rm{mH}}$, in most cases. Future work that resolves the relevant gas physics  \citep[see for example,][]{Miranda_2016,Tang_2017,Moody_2019,Munoz_2019,Munoz_2020}, and follows the evolution of BBHs down to merger boundaries of 0.1--0.01~$R_{\rm{mH}}$ is needed to further probe the poorly understood evolution of BBHs in gas disks.

\begin{table}[h]
	\centering
	\caption{The number of BBHs orbiting around their centers of mass in the prograde or retrograde direction with respect to the orbit of the disk for each run. The values for the runs shown in italics were determined only through conservation of angular momentum.}
	\label{table:pvr}
	\begin{tabular}{lccr}
		\hline
		Run & Prograde & Retrograde & Unknown\\
		\hline
		F10 & 3 & 5 & 0\\
		F20 & 3 & 4 & 1\\
        F30 & 0 & 8 & 0\\
        \em{LMA} & \em{1} & \em{5} & \em{NA}\\
        \em{LMB} & \em{2} & \em{3} & \em{NA}\\
        HMA & 1 & 4 & 1\\
        \em{HMB} & \em{5} &\em{3} & \em{NA}\\
        \hline
        Overall & 15 & 32 & 2
	\end{tabular}
\end{table}

\section{Discussion}
\label{sec:discuss}

Our models that incorporate sBHs drifting in from beyond 1000~AU every 100~kyr build up over-massive sBHs of roughly 100~M$_{\rm{\sun}}$ through hierarchical mergers resembling common aLIGO detections in less than a megayear. Incoming sBHs frequently form BBHs with sBHs on resonant orbits near the migration trap. Since resonant orbiters near the trap typically have masses in the range $20-80$~M$_{\rm{\sun}}$, BBHs formed by in-migrating ($\sim 10$~M$_{\rm{\sun}}$) sBHs with this population will tend to be asymmetric in mass ratio ($q \sim $0.1--0.5).  The ability of our model to produce asymmetric mass BBH mergers makes the environment of a migration trap in an AGN disk a promising location for the recent aLIGO merger GW190412 \citep{ligo_20}. In addition, the 60-80~M$_{\rm{\sun}}$ resonant orbiters formed in our model are similar in mass to the component masses of GW190521 \citep{LIGO_2020}, making the environment of the migration trap a promising location for GW190521 as well.

Over 10~Myr our incoming sBH models build up IMBHs of nearly 1000~M$_{\rm{\sun}}$. Our simulations produce such massive black holes because incoming sBHs are effective at perturbing other sBHs out of resonant or trojan orbits, leading nearly all sBHs in each simulation to form BBHs with the most massive black hole, located in the migration trap, over time. Therefore the inward migration of sBHs on orbits initially beyond 1000~AU and the grinding down of sBHs on inclined orbits by the gas disk make our model an even more efficient merger channel for aLIGO and Virgo detections than the scenario described in Paper I, where we only considered pre-existing objects in the inner disk. 

Our sBH replenishment rate is a loose estimate based on the most current calculations of grind down and migration rates in AGN disks \citep[Paper I,][]{FAbj_2020}, and meant to illustrate the effect incoming sBHs could have on the evolution of sBHs in the inner disk. If our sBH replenishment rate were slower or faster the periodic nature of the mergers of incoming sBHs with the outer resonant orbiter and the merger of that resonant orbiter with the sBH in the migration trap, would likely be more or less consistent, respectively. The periodicity would be less consistent with a faster replenishment rate because incoming sBHs would have the chance to migrate in before other sBHs merged more often, leading to more of the rapid back-to-back mergers described in \S \ref{sec:results:body_create}. Changing the replenishment rate of the incoming sBHs would also have an effect on the final mass of the sBH in the migration trap. A faster rate could potentially lead to an even more massive black hole in the trap, while a slower rate would lead to a less massive black hole in the trap, because there would be greater or fewer sBHs to contribute to its mass over the lifetime of the disk, respectively. However, the qualitative result, that incoming orbiters disrupt resonant convoys, and lead to additional mergers, asymmetric mass BBH mergers, and the build-up of even more massive blacks holes would not change.

We caution that our results come from using the static co-rotational torques derived by \citet{paardekooper2010}, but the hundreds of solar mass black holes that form in these runs would likely be subject to a dynamic co-rotation torque that could lead to different orbital evolution for these sBHs \citep[see Appendix A, Paper I,][]{Paardekooper_2014}.  Additionally IMBHs may also become massive enough to open a gap in the disk and therefore be subject to Type II migration \citep{lin_1986}.

In addition, we have only modeled migration in one AGN disk model, with one SMBH mass. Different disk models will have different surface densities, which will affect migration rates. The mass of the SMBH will also have an effect on the surface density of the disk. However, migration traps should occur in any disk where there is a rapid change in the surface density gradient \citep{bellovary}. Such rapid changes are likely to occur in most actual disks, since radiation pressure is expected to inflate the inner disk. \citet{Tagawa_2020} used a different AGN disk model than used in this paper, the \citet{thompson} model, which has a lower surface density than the \citet{Sirko:2003aa} disk model. The lower surface density would lead to slower migration rates, and \citet{Tagawa_2020} found that although the BBH formation rate could still be high, migration did not play as strong a role in BBH formation. However, their models focused on the region well beyond the migration trap of a \citet{thompson} AGN disk, where regardless of model parameters migration would not be as strong. \citet{Tagawa_2020} also did a parameter study of the formation of BBHs in a \citet{thompson} disk for a few different SMBH masses. They found that the rate of BBH mergers was dominated by M$_{\rm{SMBH}}=10^{7-8}$~M$_{\rm{\sun}}$. This result supports our choice to continue to focus on a model with an $10^8$~M$_{\rm{\sun}}$ SMBH. Further work examining the role of SMBH mass and AGN disk structure is needed to fully understand
the importance of the AGN disk merger channel.

With these caveats in mind, our models provide a mechanism for building 1000~M$_{\rm{\sun}}$ IMBHs that may be detectable by the Laser Interferometer Space Antenna (LISA). Typically, the most massive black hole is in the migration trap and will form a BBH with any black hole over 80~M$_{\rm{\sun}}$. The merger of an 80~M$_{\rm{\sun}}$ black hole and a black hole of several 100~M$_{\rm{\sun}}$ should be detectable first by LISA, if relatively nearby (out to approximately 0.48~Gpc at a signal-to-noise ratio of 10), and then at merger by aLIGO \citep{Flanagan_1998,Miller_2002}. At binary separations detectable with LISA, mechanisms that harden binaries (e.g.\ gas hardening or tertiary encounters) could be testable. 

Perhaps one such merger has already been detected by aLIGO/Virgo. Recent detection GW190521 resembles the merger of a resonant orbiter with the black hole in the migration trap near the beginning of an AGN disk's  merger history before the mass of the black hole in the trap builds up. For example, the first merger of a resonant orbiter with the sBH in the migration trap in the F10 run (see Figure~\ref{fig:F10hist}) is between an 80~M$_{\rm{\sun}}$ and a 90~M$_{\rm{\sun}}$ sBH.

Occasionally in our simulations BBHs with small mass ratios and total masses of around 500~M$_{\rm{\sun}}$ form. Such BBHs could be detectable during the inspiral phase by LISA out to approximately 3~Gpc with a signal-to-noise ratio of 10. In addition, if the roughly 1000~M$_{\rm{\sun}}$ IMBHs we are forming in our simulations undergo Type II migration inward, they could form SMBH-IMBH binaries. These SMBH-IMBH binaries would lead to extreme-mass ratio inspirals (EMRIs), because the mass ratio would roughly be 1:$10^5$ for $M_{\rm SMBH} = 10^8$~M$_{\rm{\sun}}$, and the mass ratio limit for EMRIs is 1:$10^4$ \citep{Amaro_Seoane_2007}. The EMRIs of these IMBHs should be detectable by LISA, depending on a variety of properties of the system.

The rate of mergers in an AGN disk was parameterized by \citet{mckernan19} as
\begin{equation}
    \label{eq:imbh_rate}
    \begin{multlined}
R = 12 \mbox{ Gpc}^{-3}\mbox{
  yr}^{-1}\frac{N_{\rm GN}}{0.006\mbox{ Mpc}^{-3}}\frac{N_{\rm BH}}{\num{2e4}}\frac{f_{\rm AGN}}{0.1} \\
   \mbox{X } \frac{f_{\rm d}}{0.1}\frac{f_{\rm b}}{0.1}\frac{\epsilon}{1}\left(\frac{\tau_{\rm AGN}}{10
  \mbox{ Myr}}\right)^{-1},
  \end{multlined}
\end{equation}
where $N_{\rm GN}$ is the average number density of galactic nuclei in the Universe, $N_{\rm BH}$ is the number of sBHs in an AGN disk, $f_{\rm AGN}$ is the fraction of galactic nuclei with AGN that last for time $\tau_{\rm AGN}$, $f_{\rm d}$ is the fraction of sBHs that end up in the AGN disk, $f_{\rm b}$ is the fraction of sBHs that form binaries, and $\epsilon$ represents the fractional change in $N_{\rm BH}$ over one full AGN duty cycle. To calculate the rate of mergers involving an IMBH (a black hole greater than 100~M$_{\rm{\sun}}$) in the inner AGN disk of our models we take $N_{\rm{BH}}f_{\rm{b}} = 17$, i.e. the number of binaries formed that involve an IMBH averaged over our different models. The resulting rate is 0.1~Gpc$^{-3}$~yr$^{-1}$, which is surprisingly similar to the rate estimated in \citet{LIGO_2020} from the detection of GW190521, 0.13~Gpc$^{-3}$~yr$^{-1}$.

For comparison, \cite{Fragione_2018a} found that in their simulations of GCs the rate of IMBH-sBH mergers was roughly 1000~Gpc$^{-3}$~yr$^{-1}$ at $z \approx 2$ and 1--10~Gpc$^{-3}$~yr$^{-1}$ at $z\approx 0$ \citep[see also,][for rates of tidal disruption events involving IMBHs]{Fragione_2018}. They also found that the merger rates are dominated by IMBHs with masses between 10$^3$ and 10$^4$~M$_{\rm{\sun}}$, mostly higher than the IMBHs formed in our models. Predicted IMBH-sBH merger rates are much higher in GCs than in our models at higher redshifts during the peak of GC formation. At lower redshifts the different rates are off by one to two orders of magnitude, but could potentially be more similar if there is IMBH formation at radii beyond the inner disk simulated here. In addition, a distinction between the two models is that IMBH-sBH mergers result from a central IMBH already present in the GC, whereas our models build-up IMBHs from sBHs. As a result, unlike in \citet{Fragione_2018a}, our rates are dominated by binaries with less massive IMBHs, suggesting that at low redshift, a low mass IMBH merger is more likely to occur in an AGN disk.

In our simulations, nearly 100\% of sBHs that are present in the vicinity of the migration trap merge over the lifetime of the disk. If this high binary fraction holds throughout the AGN disk, setting $f_{\rm{b}}$=1 in Equation \ref{eq:imbh_rate} would give a high merger rate of 120~Gpc$^{-3}$~yr$^{-1}$. If we want to calculate the BBH merger rate in the inner 1000~AU that we have modeled here alone, then the rate would be 0.66~Gpc$^{-3}$~yr$^{-1}$. However, this rate comes from setting $N_{\rm{BH}}$=110, which is a direct result of our model assumptions. Mergers of BBHs with mass ratios less than 0.5 (like GW190814 \citealt{ligo_20}) make up a significant fraction of the mergers in our models, with a merger rate of $\sim$ 0.3~Gpc$^{-3}$~yr$^{-1}$. Finally, our models suggest that if an IMBH in the migration trap were to undergo Type II migration once per the lifetime of an AGN disk, the resulting IMBH-SMBH merger rate would be 0.006~Gpc$^{-3}$~yr$^{-1}$.

The distributions of BBHs shown in Figures \ref{fig:smadist} and \ref{fig:2Dhist} are consistent with aLIGO and Virgo detections, in that the most common type of BBH merger (class (1)) involves two roughly even mass ratio BBHs that form binaries with total masses between 20 and 60~M$_{\rm \sun}$. Interestingly, class (2) of our merger channel is also efficient at producing uneven mass ratio mergers, such as GW190412 \citep{ligo_20} and sBHs with masses around 60--80~M$_{\rm \sun}$, similar to the component masses of the sBHs of recent aLIGO/Virgo detection GW190521 \citep{LIGO_2020}.  Future observations with aLIGO and Virgo and also LISA of IMBHs will help determine whether our distribution of sBH-IMBH mergers is consistent or not.

A majority of aLIGO and Virgo detections have low values for the dimensionless aligned spin of merged sBHs,
\begin{equation}
	\chi_{\rm{eff}}=\frac{\bm{S}_{\rm{1}}/M_{\rm{1}} + \bm{S}_{\rm{2}}/M_{\rm{2}}}{M_{\rm{1}}+M_{\rm{2}}} \cdot \bm{\hat{L}},
\end{equation}
where $\bm{S}_{\rm{i}}$ and $M_{\rm{i}}$ are the spin and mass, respectively, of each sBH in the merged BBH, and $\bm{\hat{L}}$ is the unit vector in the direction of the binary angular momentum. Therefore if BBH mergers occurring in AGN disks are producing a significant fraction of aLIGO and Virgo detections, the sBHs produced by our model must either have low spins or mis-aligned spins. Monte Carlo simulations performed by \citet{mckernan18} predict that for initial populations with either randomly distributed spins or near zero spins, sBHs formed from BBHs with near equal mass ratios in an AGN disk will have spins distributed around 0.7 or -0.7 depending on if the BBH orbits in the prograde or retrograde direction, respectively \citep[see also][]{Hofmann_2016,Tichy_2008,Varma_2019}. Therefore, because BBHs produced in our model are roughly evenly distributed between prograde and retrograde, second stage BBHs formed from the products of these earlier BBH mergers will often have mis-aligned spins.

Our model could also produce low spin sBHs from the mergers of BBHs orbiting around their centers of mass in the retrograde direction relative to their orbit around the SMBH, which appears to be the preferred direction of orbit in our simulations. The sBHs produced from these mergers will have spins distributed around $a \sim -0.7$ which will evolve towards low spin due to gas accretion at an average rate of $(\tau_{\rm{AGN}}/40~\rm{Myr})(\dot{\textit{m}}/\dot{M}_{\rm{Edd}})$, where $\tau_{\rm{AGN}}$ is the AGN lifetime and $\dot{\textit{m}}/\dot{M}_{\rm{Edd}}$ is the average gas accretion rate as a fraction of the Eddington rate \citep{mckernan18}. At sub-Eddington accretion rates these sBHs would spin down to low spin values on timescales on the order of the lifetime of the disk, $\mathcal{O}$(10~Myr). However, at super-Eddington accretion rates, it would be possible for $a \sim -0.7$ sBHs to spin down in a few megayears, while sBHs with spins $a \sim +0.7$ should be spun up to maximal spins on similar timescales. Thus, the slight excess of BBHs with retrograde orbits in our simulations could also produce sBHs with low spins, and super-Eddington accretion onto sBH embedded in AGN disks is an important subject for future investigation.

Additionally, even with our conservative estimate of an even distribution of prograde and retrograde binaries, it will still be more common to have sBHs with spins distributed around -0.7, because the timescale for retrograde BBHs to merge is a factor of 5 to 10 shorter than the timescale for prograde BBHs to merge \citep{baruteau2011}. Therefore, a roughly even ratio of prograde to retrograde BBHs would lead to both the formation of BBHs having mis-aligned spins, and an over-abundance of sBHs that would evolve towards low spins. \cite{Rodriguez_2018} found that in dense star clusters, such as GCs, sBHs will have anti-aligned spins after the first generation of mergers. Similarly to here, the implication is that hierarchical second or third generation mergers would then have low values of $\chi_{\rm{eff}}$. \cite{Fragione_2020a} likewise found that the distribution of $\chi_{\rm{eff}}$ of BBHs in the sBH triples scenario is distributed around zero. Conversely, parameter studies of isolated field binaries have struggled to produce either low values of $\chi_{\rm{eff}}$, or anti-aligned spins that would lead to low values of $\chi_{\rm{eff}}$ through next generation mergers \citep{Vitale_2017,Gerosa_2018}.

The models presented in this paper help deepen our understanding of the AGN disk merger channel for aLIGO and Virgo. Periodically adding sBHs from the outer disk into our simulations of the inner disk within $10^{3}$~AU increases both the number of high mass BBHs and the number of mergers in the vicinity of the migration trap. These mergers in the vicinity of the migration trap will occasionally be asymmetric in mass, resembling the recent aLIGO detection GW190412 \citep{ligo_20}, and produce 60--80~M$_{\rm\sun}$ sBHs like the component masses of GW190521 \citep{LIGO_2020}. As a result, not only can our models produce the BBH mergers that are common in aLIGO/Virgo detections, but they can also produce the more unusual recent detections. Additionally, our models illustrate that BBHs that merge in AGN disks will form with both prograde and retrograde spins. The mergers of the secondary sBHs formed from these mergers should have the low $\chi_{\rm{eff}}$ observed in aLIGO detections. Future work is needed to better understand the influence of different torque models on our simulations. Following the evolution of these BBHs down to within 0.1--0.01 $R_{\rm{mH}}$ and incorporating realistic gas physics are important to better understand how BBHs traverse the LISA detection band and enter the aLIGO detection band. Finally, as the orbital evolution of retrograde orbiters becomes better understood \citeg{FAbj_2020,secunda_2020b} simulations should incorporate sBHs orbiting in the retrograde direction as well.

\acknowledgments
A.S. would like to thank Charles Emmett Maher for useful conversations. A.S. is supported by a fellowship from the Helen Gurley Brown Revocable Trust and the NSF Graduate Research Fellowship Program under Grant No.\ DGE-1656466.  NWCL gratefully acknowledges the support of a Fondecyt Iniciacion grant \#11180005.  JMB is supported by NSF award AST-1812642. M-MML is partly supported by NSF Grant AST18-15461. ZsS is supported by NRDIO (NKFIH) grant K-119993.
KESF \& BM are supported by NSF award AST-1831412 and Simons Foundation award 533845.
\bibliographystyle{aasjournal}

\bibliography{bh_bib.bib}

\appendix

\section{A Note on Torque}
\label{A}

Here, as in Paper I, we use the unsaturated torque model from \citet{paardekooper2010}, which only considered fully unsaturated, static torques. Saturation implies that the gradient of angular momentum across the co-rotation region that drives torques from gas on horseshoe orbits has been erased, so that the gas no longer gives or takes angular momentum from the embedded orbiter. Because angular momentum from co-rotating gas is a finite resource, if no additional gas is added, the torque saturates. The torque remains unsaturated if a mechanism such as turbulent diffusion can transport fresh gas to the co-rotation region that can be tapped by the orbiter. Including saturation, as in \citet{Paardekooper_2011}, has been shown to lead to more rapid inward migration for orbiters outside a narrow mass range, and also introduce a mass dependency to both the location and existence of migration traps \citep{Hellary_2011,Coleman_2014,Dittkrist_2014}. Additionally, \cite{Paardekooper_2014} and \cite{Pierens_2015} found that co-rotation torques on orbiters in disks can often be dynamic, which can act to stall inward migration and boost outward migration. 

However, unsaturated, static torques are a reasonable approximation for the torques present in the inner radiative region of an AGN disk, which should be turbulent and have a high viscosity. When viscosity is present, it serves to desaturate the corotation torque by exchanging angular momentum between the co-rotation region and the rest of the disk \cite{Kley_2012}. \cite{Nelson_2004}, \cite{baruteau_lin}, \cite{Uribe_2011}, and \cite{baruteau2011} all found that co-rotational torques in turbulent disks are subject to stochastic turbulent fluctuations that keep the co-rotational torque unsaturated even in locally isothermal simulations. More recent simulations from \cite{Guilet_2013}, \cite{Uribe_2015}, and \cite{Comins_2016} corroborate these results, although because the width of the co-rotational region for smaller objects has yet to be resolved, it is possible for saturation to occur if turbulent fluctuations are strong enough to wipe out the horseshoe turns of these smaller orbiters. 

Dynamical co-rotation torques are also likely not to play a large role in the migration of orbiters in an AGN disk, with a \cite{shakura_sunyaev} viscosity parameter $\alpha \lesssim 10^{-2}$, as long as their masses remain below $\sim$30~M$_{\rm{\sun}}$ \citesee{Paardekooper_2014}. We therefore limit the initial mass of an orbiter in our simulations to 30~M$_{\rm{\sun}}$. However, the masses of orbiters in our incoming black holes run, described in Sections \ref{sec:diff_runs:body_create} and \ref{sec:results:body_create}, can rapidly build up to over 100~M$_{\rm{\sun}}$. These massive orbiters, which are typically formed close to the migration trap, may end up with boosted outward migration. We discuss these massive orbiters further in \S \ref{sec:discuss}.

Finally, \cite{Benitez_2015} found that a heating torque results from a proto-planet's accretion luminosity. Originally, heating torque was found to counteract inward migration, but even more importantly heating leads to significant inclination and eccentricity pumping \citep{Masset_2017,Eklund_2017}. The heating torque resulting from sBHs should be quite different than the torque resulting from planets, as sBHs do not have a surface to heat via accretion shocks. Within the Hill sphere, sBHs will be surrounded by accretion disks if local gas is optically thin, or radiation supported envelopes if local gas is optically thick. In either case, the luminosity can heat up the surrounding AGN disk gas, in analogy with the mechanism found by \citet{Benitez_2015}. Recent work by \cite{Hankla_2020}, which assumed thermal feedback and Eddington-limited accretion with no mechanical feedback, showed that a heating torque does lead to migration. They even showed that heating torque can be the dominant physical effect in regions of an AGN disk where optical depth is moderate ($\lesssim 300$). The optical depth in the inner region of a \citet{Sirko:2003aa} AGN disk remains above this threshold, although the optical depth of the outer disk (beyond $10^4$~R$_{\rm s}$) is well below this threshold (see Figure~\ref{fig:sirkoM}). We therefore defer further investigation of the important question of sBH feedback and its effect on the dynamics of a sBH in a gas disk to future work.

\end{document}